\documentclass[12pt]{article}

\usepackage[T1]{fontenc}
\usepackage{lmodern}
\usepackage[utf8]{inputenc}
\usepackage[american]{babel}

\usepackage{geometry}
\usepackage{setspace}
\usepackage{microtype}
\usepackage[section]{placeins}

\usepackage{amsmath, amssymb}
\usepackage{graphicx}
\usepackage{booktabs}
\usepackage{threeparttable}
\usepackage{array}
\usepackage{tabularx}
\usepackage{longtable}
\usepackage{caption}
\usepackage{float}
\usepackage{ragged2e}
\usepackage{siunitx}
\usepackage{makecell}

\graphicspath{{figures/}{./}}
\setkeys{Gin}{width=\linewidth,keepaspectratio}

\usepackage{xcolor}

\newcolumntype{Y}{>{\RaggedRight\arraybackslash}X}

\usepackage{authblk}


\usepackage{csquotes}
\usepackage[style=apa, backend=biber, maxcitenames=2, uniquelist=false]{biblatex}
\DeclareLanguageMapping{american}{american-apa}
\usepackage{xurl}
\usepackage[unicode,breaklinks=true]{hyperref}
\hypersetup{
  colorlinks=true,
  urlcolor=blue,
  linkcolor=black,
  citecolor=black
}

\usepackage{tikz}
\usetikzlibrary{positioning, fit, backgrounds, arrows.meta}

\title{A Multi-Agent Orchestration Framework for Venture Capital Due Diligence}
\author{Grigorios Alexandrou}
\author{Katerina Pramatari}
\affil{Department of Management Science and Technology,
Athens University of Economics and Business (AUEB), Athens, Greece}
\date{\today}

\begin{document}
\maketitle

\begin{abstract}
  We present a fully automated multi-agent framework for corporate due diligence and market
  analysis in venture capital. The system runs on an event-driven orchestration architecture,
  combining Large Language Models (LLMs) with real-time web retrieval to synthesize unstructured
  data into structured investment intelligence. A central technical contribution is a programmatic
  extraction pipeline that reverse-engineers the frontend-to-backend communication of the Greek
  Business Registry ($\Gamma$.E.MH.), querying dynamic endpoints to retrieve official financial
  filings that are then parsed using a layout-aware OCR extractor. A structural fallback mechanism
  explicitly flags data absence rather than generating unverified figures, directly targeting
  hallucination in financial contexts. All workflow artifacts are publicly available to support
  replication.
\end{abstract}


\section{Introduction}
\label{sec:intro}

Evaluating prospective investments and monitoring portfolio companies demands the synthesis of
fragmented information across incompatible sources: unstructured web data, real-time news,
competitive analyses, and official financial registries that are often locked behind
administrative access controls. Manual desk research, the standard approach, introduces human
error, cognitive bias, and delays that compound across a portfolio
\parencite{gompers2016private, kahneman2011thinking}.

LLMs offer powerful capabilities for natural language understanding and data synthesis
\parencite{bommasani2021opportunities}, but deploying them for financial due diligence exposes
three structural problems. First, training data cut-offs leave models blind to current market
signals. Second, LLMs generate plausible but factually incorrect information---a failure mode that
financial decision-making cannot tolerate \parencite{ji2023survey}. Third, base LLMs lack the
ability to autonomously query external databases, invoke APIs, or parse scanned financial
documents; frameworks that enable tool use and inter-agent communication are required to bridge
this gap \parencite{wu2023autogen}.

We address these problems with an event-driven multi-agent orchestration pipeline. Rather than a
single LLM prompt, the system decomposes due diligence into specialized sub-tasks handled by
distinct AI agents \parencite{xi2023rise}. Real-time search APIs supply current market
intelligence; a custom extraction module programmatically retrieves and parses official financial
filings from the Greek Business Registry ($\Gamma$.E.MH.).

This paper makes three contributions:
\begin{enumerate}
  \item A modular, multi-agent architecture that automates end-to-end corporate research from a
        single analyst trigger to a structured HTML report delivered by email.
  \item A programmatic financial extraction pipeline that queries hidden $\Gamma$.E.MH.\ registry
        endpoints, retrieves raw PDF documents, and applies layout-aware OCR to structure
        financial data with preserved source citations.
  \item A structural fallback mechanism that routes to third-party commercial financial databases when
        registry data is unavailable, replacing potential LLM hallucinations with explicit,
        auditable data gaps.
\end{enumerate}

\section{Background and Related Work}
\label{sec:background}

\subsection{Multi-Agent Frameworks}

The field has evolved from single-prompt LLM interactions toward agentic workflows, where multiple
LLM-powered agents each carry distinct roles, system prompts, and tool-use capabilities
\parencite{xi2023rise}. Chaining agents allows each one to pass its verified output to the next,
constructing reasoning chains that exceed the capacity of a single model call. Multi-step
analytical problems requiring precision and iterative correction are the natural application
domain for these architectures.

AutoGen operationalizes this paradigm through a conversational multi-agent framework in which
agents negotiate task decomposition and tool invocation dynamically \parencite{wu2023autogen}.
MetaGPT takes a complementary approach, assigning software-engineering roles---product manager,
architect, engineer---to distinct agents coordinated via structured output schemas
\parencite{hong2023metagpt}. LangGraph provides a graph-based state-machine abstraction for agent
orchestration, enabling conditional branching and cycle detection in complex pipelines
\parencite{langgraph2024}. Our system follows the same DAG-structured philosophy but is
implemented on n8n \parencite{n8n2024}, a low-code event-driven platform that exposes HTTP,
conditional routing, and AI nodes without requiring custom agent code, substantially lowering the
engineering barrier for non-technical investment teams.

\subsection{LLMs for Financial Analysis}

Applying agentic systems to finance is natural: general-purpose LLMs struggle with domain-specific
vocabulary and numerical precision \parencite{yang2023fingpt}. BloombergGPT demonstrated that
domain-adapted pre-training yields measurable improvements on financial NLP benchmarks
\parencite{wu2023bloomberggpt}. More recent multi-agent architectures have targeted specific
financial tasks: MARAG-Fin distributes analytical subtasks across agents to improve trading
decision accuracy and reduce hallucinations in volatile markets \parencite{maragfin2025}, while
QuantAgents structures investment analysis as a simulated trading environment coordinated by
specialist agents \parencite{li2025quantagents}. FinAgent extends this line by incorporating
multimodal inputs---price charts, news sentiment, and fundamental data---into a unified agent
pipeline \parencite{zhang2024finagent}. Our work is distinguished from this body of literature by
its focus on pre-investment due diligence rather than post-investment trading, and by its
integration with an official national corporate registry that provides auditable, ground-truth
financial data rather than market signals.

\subsection{Retrieval-Augmented Generation and Real-Time Grounding}

Dynamic information retrieval addresses the static knowledge problem: grounding model outputs in
live search results and external data streams allows systems to bypass training cut-offs entirely
\parencite{lewis2020retrieval}. For corporate intelligence, Open-Source Intelligence (OSINT)
provides current market sizing, funding rounds, and competitor movements.
Retrieval-Augmented Generation makes that intelligence auditable by linking outputs to primary
sources \parencite{gao2023retrieval}.

\subsection{Document Understanding and OCR}

Automated document processing extends retrieval into non-digitized corporate records.
Layout-aware parsers handle balance sheets and income statements without the structural loss
typical of legacy OCR engines \parencite{paruchuri2025marker}. Combined with real-time web
retrieval, these tools provide a path from fragmented market signals to formal regulatory filings
within a single pipeline---the integration this work demonstrates.

\section{System Architecture}
\label{sec:architecture}

The system runs on an event-driven automation platform \parencite{n8n2024}, structured as a
Directed Acyclic Graph (DAG) of processing nodes. A custom HTML form serves as the entry point
for investment teams: analysts select a target portfolio company and trigger the full research
workflow with a single click, with the low-code backend hidden from view.

\begin{figure}[htbp]
  \centering
  \begin{tikzpicture}[
      arr/.style={->, >=stealth, semithick, black!55},
      lbl/.style={draw=none, fill=none, font=\scriptsize\itshape,
          black!65, inner sep=1pt},
      boxnode/.style={rectangle, rounded corners=4pt, draw=gray!55,
          fill=white, align=center, font=\small, inner sep=6pt},
    ]

    \node[boxnode, text width=6cm] at (0, 0) (trigger)
    {\textbf{HTML Form Trigger}\\[2pt]
    {\footnotesize Analyst selects portfolio company}};

    \node[boxnode, text width=12cm] at (0, -2.2) (intake)
    {\textbf{3.1 \; Data Intake \& Context Initialization}\\[3pt]
    {\footnotesize Webhook $\;\rightarrow\;$ JSON Company DB
    $\;\rightarrow\;$ \textbf{AI Context Agent}}};

    \node[boxnode, text width=2.5cm, font=\footnotesize] at (-6.5, -5.5) (mapper)
    {\textbf{AI Source Mapper}\\[2pt]{\scriptsize Identifies sources}};
    \node[boxnode, text width=2.2cm, font=\footnotesize] at (-3.5, -5.5) (sector)
    {\textbf{AI Sector}\\[2pt]{\scriptsize Market sizing}};
    \node[boxnode, text width=3.5cm, font=\footnotesize] at ( 0.0, -5.5) (competition)
    {\textbf{AI Competition}\\[2pt]{\scriptsize Competitor landscape}};
    \node[boxnode, text width=2.2cm, font=\footnotesize] at ( 3.5, -5.5) (news)
    {\textbf{AI News}\\[2pt]{\scriptsize Recent news}};
    \node[boxnode, text width=2.5cm, font=\footnotesize] at ( 6.5, -5.5) (signals)
    {\textbf{AI Signals}\\[2pt]{\scriptsize Investment signals}};

    \begin{scope}[on background layer]
      \node[draw=gray!50, fill=white, dashed, rounded corners=5pt,
        inner sep=17pt,
        fit=(mapper)(sector)(competition)(news)(signals),
        label={[draw=none, fill=none, font=\small\bfseries, black!60,
                anchor=south, yshift=2pt]above:%
            3.2 \; Market \& Competitive Intelligence
          }]
      (group32){};
    \end{scope}

    \node[boxnode, text width=3.5cm, font=\footnotesize] at (-5.0, -9.5) (gemh)
    {\textbf{$\Gamma$.E.MH.\ Module}\\[2pt]
    {\scriptsize Endpoint query, OCR \& Financial Analysis}};
    \node[boxnode, text width=3.5cm, font=\footnotesize] at ( 0.0,  -9.5) (router)
    {\textbf{Exists on $\Gamma$.E.MH.?}\\[2pt]
    {\scriptsize Conditional router}};
    \node[boxnode, text width=3.5cm, font=\footnotesize] at ( 5.0, -9.5) (altfin)
    {\textbf{Alt.\ Financials}\\[2pt]
    {\scriptsize Crunchbase / Dealroom\\
    \textit{or} ``Not Found'' flag}};

    \begin{scope}[on background layer]
      \node[draw=gray!50, fill=white, dashed, rounded corners=5pt,
        inner sep=12pt,
        fit=(gemh)(router)(altfin),
        label={[draw=none, fill=none, font=\small\bfseries, black!60,
                anchor=south, yshift=2pt]above:%
            3.3 \; Financial Data Retrieval \& Fallback}]
      (group33){};
    \end{scope}

    \node[boxnode, text width=16cm] at (0, -12.5) (synthesis)
    {\textbf{3.4 \; Synthesis \& Output Generation}\\[3pt]
    {\footnotesize \textbf{AI Researcher} $\;\rightarrow\;$
    \textbf{AI Analyst} $\;\rightarrow\;$
    \textbf{AI Overall Company Info} $\;\rightarrow\;$
    HTML report $\;\rightarrow\;$ Gmail}};

    \draw[arr] (trigger)       -- (intake);
    \draw[arr] (intake)        -- (group32.north);
    \draw[arr] (group32.south) -- (router.north);
    \draw[arr] (router.west)   -- node[lbl, above]{Yes} (gemh.east);
    \draw[arr] (router.east)   -- node[lbl, above]{No}  (altfin.west);
    \draw[arr] (group33.south) -- (synthesis.north);

  \end{tikzpicture}
  \caption{High-level overview of the event-driven multi-agent orchestration
    architecture, structured according to the four pipeline stages described
    in Sections~\ref{sec:intake}--\ref{sec:output}.}
  \label{fig:architecture}
\end{figure}
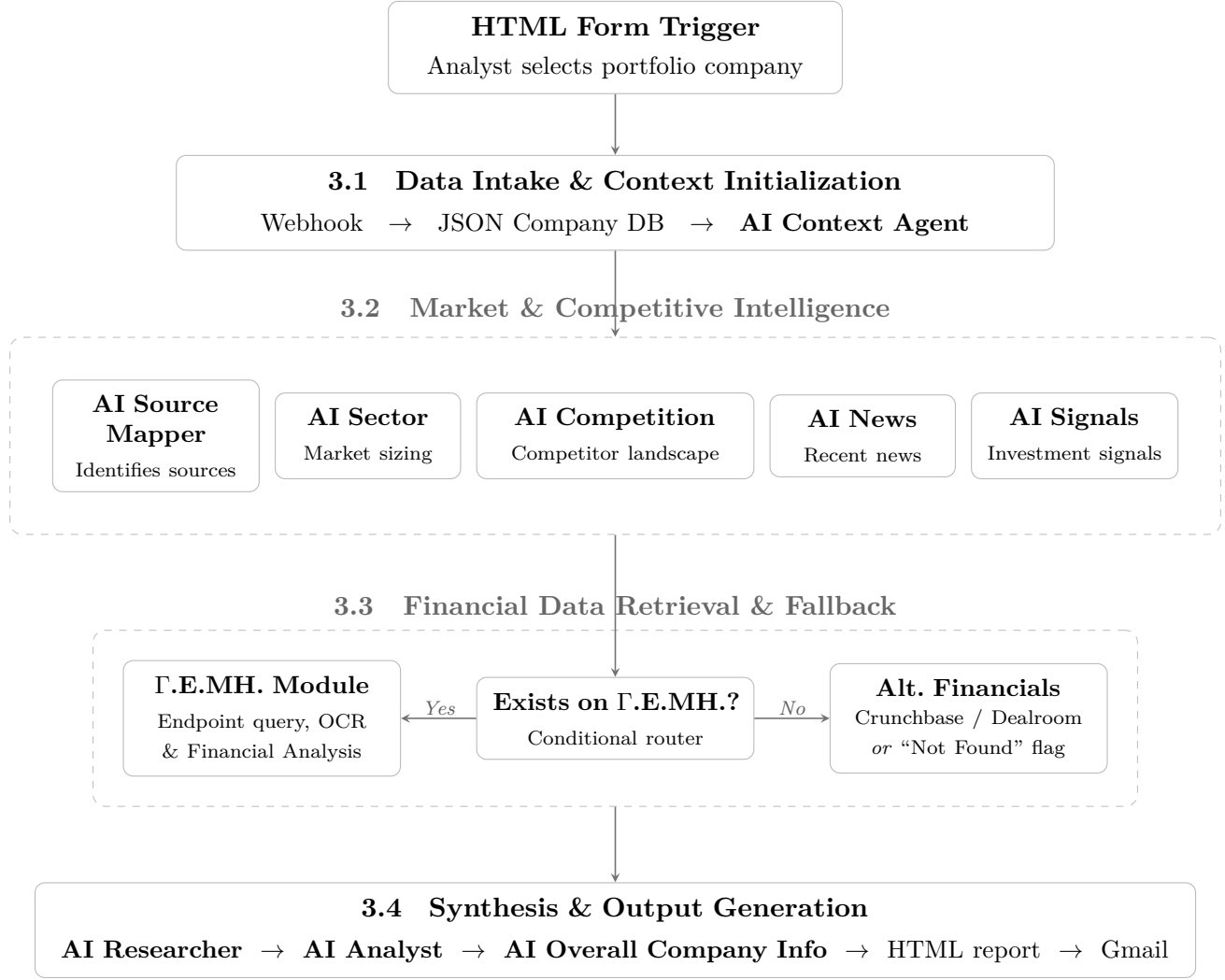

\subsection{Data Intake and Context Initialization}
\label{sec:intake}
On submission, a Webhook node captures the payload. A transformation node maps the selected
company to a pre-scraped JSON database of baseline attributes: founders, sector, initial
investment year, headquarters, and registration numbers. The \textbf{AI Context Agent} takes this
structured input and produces a compact company profile that anchors all downstream research nodes.

\subsection{Market and Competitive Intelligence Modules}
Five specialized agents query the Perplexity Sonar Deep Research API
\parencite{perplexity2024sonar} for real-time market intelligence. The \textbf{AI Source Mapper}
runs first, identifying the most reliable sector-specific portals, databases, and news aggregators
for the target company. The \textbf{AI Sector} and \textbf{AI Competition} agents then run in
parallel against those sources, extracting market sizing, macro-environmental trends, and a
competitive landscape that categorizes competitors as direct, adjacent, or niche innovators. Once
both complete, the \textbf{AI News} and \textbf{AI Signals} agents scan for recent strategic
developments (partnerships, product launches) and extract quantifiable investment signals.

\subsection{Financial Data Retrieval and Fallback Mechanisms}
\label{sec:financials}

The most technically involved component interfaces with the Greek Business Registry
($\Gamma$.E.MH.). Although the registry provides an official API, obtaining authenticated
credentials involves administrative overhead. The framework instead reverse-engineers the portal's
frontend-to-backend communication, a method validated by prior open-source work targeting the same
registry \parencite{drakakis2025govdoc}.

Registry data and corporate announcements are served as PDF documents loaded from specific backend
endpoints. A conditional router checks for a valid registry number (\textit{arGEMI}). When one
exists, an HTTP Request node mimics browser behavior, querying those endpoints to retrieve the
JSON payload of document IDs.

Document IDs are classified into two streams: \textbf{Corporate Modifications}---covering board
changes, capital increases, and statutory modifications---and \textbf{Financial Statements},
comprising official balance sheets, income statements, and auditor reports. The pipeline fetches
raw PDFs, prioritizing the most recent fiscal years, and passes them through a layout-aware text
extractor \parencite{paruchuri2025marker}. Two summarization agents, \textbf{AI FinSummary} and
\textbf{AI ModSummary}, distill the extracted text into standardized financial metrics (Assets,
Liabilities, Revenue, EBIT) with source citations preserved.

Preventing fabricated financial data is the system's hardest constraint. When no $\Gamma$.E.MH.\
number is found---typically for companies incorporated abroad---the pipeline activates the
\textbf{Alternative Financials} agent, which queries Crunchbase, Dealroom, and equivalent
third-party commercial databases. If those sources also return nothing, the agent writes a ``Not Found'' flag
directly into the report, making the data gap explicit rather than filling it with generated
figures.

\subsection{Synthesis and Output Generation}
\label{sec:output}

Three synthesis agents combine the intelligence streams. The \textbf{AI Researcher} aggregates
sector, competition, news, signals, and financial data into a strategic research note that surfaces
recent developments and blind spots. The \textbf{AI Analyst} takes that note, produces an
Executive Summary, assigns attractiveness scores across market timing and product differentiation,
and drafts 30-to-180-day recommendations for both the fund and the startup. The \textbf{AI Overall
Company Info} agent generates a public-facing summary that founders can use to understand their
external market footprint.

A formatting node converts the JSON outputs into a styled HTML report and delivers it to the
requesting analyst through a Gmail API integration.

\section{Report Structure and Fallback Behavior}
\label{sec:report}

The final artifact of the pipeline is a structured HTML report delivered by email. Its sections
map directly to the agent layer that produced them, making the provenance of each claim traceable
to a specific intelligence stream.

\subsection{Report Anatomy}

The report opens with a \textbf{Company Overview} produced by the \textbf{AI Overall Company Info}
agent: a concise external-facing summary of the portfolio company's market position, value
proposition, and competitive footprint as reflected in publicly available signals. This section is
intentionally written to be shareable with founders as a reflection of their external presence.

The \textbf{Market Intelligence} block covers current market sizing and trajectory, macro-
regulatory dynamics, and a timeline of recent strategic events (funding announcements, product
launches, partnerships). Each claim carries an inline citation to the underlying Perplexity Sonar
source retrieved by the \textbf{AI Sector} and \textbf{AI News} agents.

The \textbf{Competitive Landscape} section, contributed by the \textbf{AI Competition} agent,
classifies identified competitors into three tiers---direct, adjacent, and niche innovators---with
funding status, activity signals, and a positioning summary relative to the portfolio company.

The \textbf{Financial Summary} presents the structured line items extracted from $\Gamma$.E.MH.\
official filings: Revenue, Total Assets, Total Liabilities, and EBIT across available fiscal
years, each cited to the source PDF and page number extracted by the \textbf{AI FinSummary} agent.
A \textbf{Corporate Events} timeline, produced by \textbf{AI ModSummary}, lists statutory changes
such as capital increases and board amendments in chronological order. When registry data is
unavailable, this block displays either a structured entry from Crunchbase or Dealroom, or an
explicit ``Not Found'' flag (see Section~\ref{sec:financials}).

The report closes with an \textbf{Analyst Assessment} from the \textbf{AI Analyst} agent:
an Executive Summary, attractiveness scores across market timing and product differentiation
dimensions, and 30-to-180-day action recommendations addressed separately to the fund and to the
portfolio company.

\subsection{Fallback Behavior in Practice}

The structural fallback is directly observable in the report output. When a company holds no valid
\textit{arGEMI}---typically because it is incorporated abroad---the Financial Summary block
displays a ``Not Found'' flag rather than any generated estimate. This flag is rendered with the
same formatting as a populated entry, making the data gap explicit and immediately visible to the
analyst. The design reflects a deliberate trade-off: a missing figure is less harmful to
investment analysis than a plausible but unverifiable one. Analysts can therefore immediately
distinguish registry-sourced figures (auditable, official) from third-party commercial
approximations (Crunchbase, Dealroom) from explicit gaps---three distinct epistemic states that the system
surfaces without conflating them.

\section{Limitations}
\label{sec:limitations}

Several constraints bound the current system. First, the financial extraction component is
specific to $\Gamma$.E.MH., limiting registry-grade data to Greek-incorporated entities; companies
registered abroad rely on third-party commercial databases with lower coverage and less structured data.
Second, the pipeline depends on three external commercial services---Perplexity Sonar, the LLM
provider, and the Marker PDF extractor---making it subject to pricing changes, rate limits, and
interface updates; a production deployment should consider self-hosted alternatives for
mission-critical components. Third, the current deployment has been applied within a single Greek venture capital fund;
generalizability across investment stages, sectors, and geographies remains to be established
through broader empirical evaluation. Fourth,
LLM inference is non-deterministic: repeated runs on identical inputs may produce different
synthesis outputs, and the variability of the generated sections is currently unquantified.
Setting the inference temperature to zero would substantially reduce this variance; logging
output hashes across repeated runs would provide an empirical measure of inter-run divergence.

\section{Data and Workflow Availability}
\label{sec:data}

All system artifacts are publicly available in a dedicated repository
\parencite{alexandrou2026due}: the n8n JSON workflow file and setup documentation. The repository
README details the required credentials and node configuration necessary for replication on a
self-hosted or cloud n8n instance.

\section{Conclusions and Future Work}
\label{sec:conclusion}

Multi-agent orchestration can automate corporate due diligence at a level of reliability
previously requiring human analysts. By combining dynamic web retrieval, reverse-engineered
endpoint querying, and OCR-based document processing, the pipeline connects unstructured market
signals to official financial compliance data in a single automated pass. Structural fallback
mechanisms prevent the system from generating unverified financial figures, replacing potential
hallucinations with explicit data gaps.

Three directions extend this work. First, integrating the pipeline with the fund's internal
portfolio database would let each report draw on proprietary deal-level data, historical
performance metrics, and analyst notes specific to each startup, making the output contextual
rather than generic. Second, adding international registries---such as UK Companies House
\parencite{companieshouse2025} and equivalent European authorities---would extend financial
extraction coverage to companies incorporated outside Greece. Third, replacing the third-party PDF
extraction service with a self-hosted open-source alternative would reduce operational costs and
remove the external dependency without degrading extraction quality.

\section*{Acknowledgments}

The authors thank the team at Uni.Fund for providing the operational context in which this system
was developed and deployed, and for their feedback on the generated reports. This work was carried
out at the Department of Management Science and Technology, Athens University of Economics and
Business (AUEB).

\printbibliography

@inproceedings{lewis2020retrieval,
  title     = {Retrieval-augmented generation for knowledge-intensive NLP tasks},
  author    = {Lewis, Patrick and Perez, Ethan and Piktus, Aleksandra and Petroni, Fabio and Karpukhin, Vladimir and Goyal, Naman and K{\"u}ttler, Heinrich and Lewis, Mike and Yih, Wen-tau and Rockt{\"a}schel, Tim and others},
  booktitle = {Advances in Neural Information Processing Systems (NeurIPS)},
  volume    = {33},
  pages     = {9459--9474},
  year      = {2020}
}

@article{xi2023rise,
  title   = {The rise and potential of large language model based agents: A survey},
  author  = {Xi, Zhiheng and Chen, Wenxiang and Guo, Xin and He, Wei and Ding, Yi and Hong, Boyang and Zhang, Ming and Wang, Junzhe and Jin, Senjie and Zhou, Enyu and others},
  journal = {arXiv preprint arXiv:2309.07864},
  year    = {2023}
}

@article{ji2023survey,
  title     = {Survey of hallucination in natural language generation},
  author    = {Ji, Ziwei and Lee, Nayeon and Frieske, Rita and Yu, Tiezheng and Su, Dan and Xu, Yan and Ishii, Etsuko and Bang, Yejin and Madotto, Andrea and Fung, Pascale},
  journal   = {ACM Computing Surveys},
  volume    = {55},
  number    = {12},
  pages     = {1--38},
  year      = {2023},
  publisher = {ACM New York, NY},
  url       = {https://arxiv.org/pdf/2202.03629}
}

@article{yang2023fingpt,
  title   = {{FinGPT}: Open-source financial large language models},
  author  = {Yang, Hongyang and Liu, Xiao-Yang and Wang, Christina Dan},
  journal = {arXiv preprint arXiv:2306.06031},
  year    = {2023}
}

@article{bommasani2021opportunities,
  title         = {On the opportunities and risks of foundation models},
  author        = {Bommasani, Rishi and Hudson, Drew A and Adeli, Ehsan and others},
  journal       = {arXiv preprint arXiv:2108.07258},
  year          = {2021},
  url           = {https://arxiv.org/pdf/2108.07258},
  eprint        = {2108.07258},
  archiveprefix = {arXiv}
}

@article{wu2023autogen,
  title   = {Autogen: Enabling next-gen llm applications via multi-agent conversation},
  author  = {Wu, Qingyun and Bansal, Gagan and Zhang, Jieyu and others},
  journal = {arXiv preprint arXiv:2308.08155},
  year    = {2023}
}

@article{gao2023retrieval,
  title         = {Retrieval-augmented generation for large language models: A survey},
  author        = {Gao, Yunfan and Xiong, Yun and Gao, Xinyu and others},
  journal       = {arXiv preprint arXiv:2312.10997},
  year          = {2023},
  url           = {https://arxiv.org/pdf/2312.10997},
  eprint        = {2312.10997},
  archiveprefix = {arXiv}
}

@article{wu2023bloomberggpt,
  title   = {BloombergGPT: A Large Language Model for Finance},
  author  = {Wu, Shijie and Irsoy, Ozan and others},
  journal = {arXiv preprint arXiv:2303.17564},
  year    = {2023}
}

@software{alexandrou2026due,
  author    = {Alexandrou, Grigorios},
  title     = {A Multi-Agent Framework for Automated Due Diligence: n8n Workflow and Documentation},
  year      = {2026},
  publisher = {GitHub},
  url       = {https://github.com/gregalexan/ai-researcher-latex/tree/main/workflow},
  howpublished = {\url{https://github.com/gregalexan/ai-researcher-latex/tree/main/workflow}}
}

@article{gompers2016private,
  title     = {What do private equity firms say they do?},
  author    = {Gompers, Paul A and Kaplan, Steven N and Mukharlyamov, Vladimir},
  journal   = {Journal of Financial Economics},
  volume    = {121},
  number    = {3},
  pages     = {449--476},
  year      = {2016},
  publisher = {Elsevier},
  url       = {https://www.hbs.edu/ris/Publication%20Files/15-081_9baffe73-8ec2-404f-9d62-ee0d825ca5b5.pdf}
}

@book{kahneman2011thinking,
  title     = {Thinking, Fast and Slow},
  author    = {Kahneman, Daniel},
  year      = {2011},
  publisher = {Farrar, Straus and Giroux},
  address   = {New York}
}

@misc{n8n2024,
  author       = {{n8n}},
  title        = {n8n: Workflow Automation Platform},
  year         = {2024},
  howpublished = {\url{https://n8n.io}},
  note         = {Accessed: 2026-03-04}
}

@misc{perplexity2024sonar,
  author       = {{Perplexity AI}},
  title        = {Sonar: Real-Time Web Search {API}},
  year         = {2024},
  howpublished = {\url{https://docs.perplexity.ai}},
  note         = {Accessed: 2026-03-04}
}

@software{paruchuri2025marker,
  author       = {Paruchuri, Vik},
  title        = {Marker: Fast, high-accuracy PDF to Markdown conversion},
  year         = {2025},
  publisher    = {GitHub},
  journal      = {GitHub repository},
  howpublished = {\url{https://github.com/datalab-to/marker}}
}

@misc{drakakis2025govdoc,
  author       = {Drakakis, Eftihis},
  title        = {Flexible GovDoc Scanner},
  year         = {2025},
  howpublished = {Google Summer of Code, Open Technologies Alliance - GFOSS},
  url          = {https://summerofcode.withgoogle.com/programs/2025/projects/dKzwxQT9}
}

@article{maragfin2025,
  title   = {MARAG-Fin: An Intelligent Multi-agent RAG-LLM Architecture Integrating Financial News Sentiment and Time Series Data for Data-driven Trading Decision-making},
  author  = {Luckianto, Marvin and Gunawan, Alexander Agung Santoso},
  journal = {International Journal of Intelligent Engineering and Systems},
  volume  = {19},
  number  = {2},
  pages   = {740},
  year    = {2026},
  doi     = {10.22266/ijies2026.0228.46}
}

@inproceedings{li2025quantagents,
  title     = {QuantAgents: Towards Multi-agent Financial System via Simulated Trading},
  author    = {Li, Xiangyu and Zeng, Yawen and Xing, Xiaofen and Xu, Jin and Xu, Xiangmin},
  booktitle = {Findings of the Association for Computational Linguistics: EMNLP 2025},
  pages     = {17438--17464},
  year      = {2025},
  publisher = {Association for Computational Linguistics}
}

@inproceedings{hong2023metagpt,
  title     = {{MetaGPT}: Meta Programming for a Multi-Agent Collaborative Framework},
  author    = {Hong, Sirui and Zhuge, Mingchen and Chen, Jonathan and Zheng, Xiawu and Cheng, Yuheng and Zhang, Ceyao and Wang, Jinlin and Wang, Zili and Yau, Steven Ka Shing and Lin, Zijuan and others},
  booktitle = {The Twelfth International Conference on Learning Representations (ICLR)},
  year      = {2024},
  url       = {https://arxiv.org/abs/2308.00352}
}

@software{langgraph2024,
  author       = {{LangChain Inc.}},
  title        = {{LangGraph}: Build Resilient Language Agents as Graphs},
  year         = {2024},
  publisher    = {GitHub},
  howpublished = {\url{https://github.com/langchain-ai/langgraph}}
}

@article{zhang2024finagent,
  title   = {{FinAgent}: A Multimodal Foundation Agent for Financial Trading:
             Data-Retrieval, Data-Mining, Policy-Generation, and Trading},
  author  = {Zhang, Wentao and Zhao, Lingxuan and Xia, Haochong and Sun, Shuo and Sun, Jiaze and Zhao, Molei and Li, Xinyi and Zhao, Yuqing and Shu, Yilei and Du, Fangyi and others},
  journal = {arXiv preprint arXiv:2402.18485},
  year    = {2024}
}

@misc{companieshouse2025,
  author       = {{UK Companies House}},
  title        = {Companies House Public Data API},
  year         = {2025},
  howpublished = {\url{https://developer.company-information.service.gov.uk/}}
}

\end{document}